\documentclass[10pt,twocolumn,showpacs,preprintnumbers,amsmath,amssymb,aps,prb,longbibliography,superscriptaddress,nofootinbib]{revtex4-2}
\usepackage{array,braket,mathtools,siunitx}

\usepackage[unicode=true,
bookmarks=true,bookmarksnumbered=true,bookmarksopen=false,
breaklinks=true,pdfborder={0 0 0},backref=false,colorlinks=true]{hyperref}
\hypersetup{citecolor={blue},urlcolor={magenta}}

\usepackage{cleveref}
\crefname{appendix}{App.}{Apps.}
\crefname{equation}{Eq.}{Eqs.}
\crefname{figure}{Fig.}{Figs.}
\crefname{table}{Tab.}{Tabs.}
\crefname{section}{Sec.}{Secs.}

\newcommand{\bR}{\mathbf{R}}

\newcommand{\bP}{\mathbf{P}}

\begin{document}
\title{Two-dimensional quantum breakdown model with Krylov subspace many-body localization}
\author{Xinyu Liu}
\affiliation{Department of Physics, California Institute of Technology,
Pasadena, California 91125, USA}
\affiliation{Department of Physics, Princeton University, Princeton, New Jersey 08544, USA}
\author{Biao Lian}
\affiliation{Department of Physics, Princeton University, Princeton, New Jersey 08544, USA}

\begin{abstract}
We propose a two-dimensional (2d) quantum breakdown model of hardcore bosons interacting with disordered spins which would be classical without the bosons. It resembles particles incident into supersaturated vapor. The model exhibits a set of subsystem symmetries, and has a strong fragmentation into Krylov subspaces in each symmetry sector. The Hamiltonian in each Krylov subspace maps to a single-particle problem in a Cayley tree-like graph. At zero disorder, the Krylov subspaces exhibit either (possible) integrable features, or quantum chaos with quantum scar states showing irregular energy and degeneracy patterns. At nonzero disorders, they enter a 2d many-body localization (MBL) phase beyond certain disorder strength $W_*$, as indicated by Poisson level spacing statistics and entanglement entropy growing as $\log t$ with time $t$. Our theoretical arguments suggest $W_*$ is finite or zero for boson number $N_b\lesssim L^\gamma/\log L$ ($1/2\le \gamma \le 1$) as system size $L\rightarrow\infty$. This gives a more stringent condition for MBL than that in the 1d quantum breakdown models. This model reveals the possibility of MBL in systems of quantum particles interacting with classical degrees of freedom.
\end{abstract}

\maketitle

\section{Introduction}

Non-thermalizing many-body quantum systems have been attracting extensive interests, which may have applications in coherent controls and storage of quantum information. Ideas for achieving non-thermalization include construction of models with quantum scars \cite{bernien2017,moudgalya2018,schecter2018,turner2018a,choi2019scar,ho2019,bull2019,lin2019,khemani2019,scherg2021,kao2021,jepsen2022,su2022scar,desaules2022,desaules2022b} or Hilbert space fragmentation \cite{Moudgalya_2021,herviou2021,sala2020,khemani2020,znidaric2013,yangzc2020,moudgalya2020,tomasi2019,buca2022,buca2023,lian2023}, and exploration of many-body localization (MBL) phases \cite{basko2006,gornyi2005,oganesyan2007,marko2008,nandkishore2015,abaninRMP2019} typically requiring disorders. Particularly, the identification of MBL remains challenging theoretically and numerically \cite{roeck2017,thiery2018,luitz2017,goihl2019,crowley2020,morningstar2022,jan2022,sels2022} even for the most-studied interacting fermion or spin models in one-dimension (1d), and is more difficult in higher dimensions. The limitation of numerical calculations urges the exploration of models with solvable limits or reducible complexity for the search of MBL.

Instructive results have been obtained by reformulating the MBL problem as a single-particle Anderson localization problem in Fock space \cite{gornyi2005,altshuler1997,berkovits1998,leyronas1999,flambaum2001,monthus2010,logan2019,roy2020b}. It was argued the geometry of Fock space is similar to the Cayley tree, in which case the criterion for Anderson localization in Cayley tree \cite{abou1973,abou1974,mirlin1997,chalker1990,savitz2019,baroni2023,sade2003,monthus2011,tikhonov2016,tikhonov2016frac,roy2020} applies. However, the approximate nature of the Cayley tree picture obstructed rigorous verifications of MBL, and quite often the large coordination numbers of the effective Cayley tree picture make it difficult to achieve MBL.

Motivated by the recently studied 1d quantum breakdown model \cite{lian2023,hu2024_boson,chen2024_breakdown,hu_gauge2024}, we propose a 2d quantum breakdown model of hardcore bosons with a breakdown interaction with the disordered spins for MBL in a 2d square lattice, in which the boson number $N_b$ is conserved. The spins would be classical and non-dynamical with a random magnetic field if there is no bosons. Such a system resembles the cloud chamber which is a celebrated particle detection device, where the bosons interacting with spins resemble particles incident into a supersaturated vapor. The model possesses a set of intriguing subsystem symmetries, while each symmetry sector further exhibits Hilbert space fragmentation into exponentially many Krylov subspaces. Each Krylov subspace has an exact mapping to a modified single-particle Anderson localization problem in a Cayley-tree like graph in Fock space with correlated disorders, where the coordination number is limited by the boson number. 

At zero disorder, we find the Krylov subspaces exhibit either integrable features or quantum chaos with (degenerate) many-body quantum scar eigenstates. In particular, we show a special sector of single boson (total boson number $N_b=1$) at zero disorder decomposes into many integrable Krylov subspaces which map to the exactly solvable tight-binding model in the Cayley tree \cite{ostilli2022}. In multi-boson sectors at zero disorder, via numerical exact diagonalization (ED), we identified intriguing Krylov subspaces showing quantum scar states with an irregular energy and degeneracy pattern, which gives rise to non-thermalizing quantum dynamics for certain initial (Fock) states.

At nonzero disorders, our ED calculation indicates evident 2d MBL beyond certain disorder strength $W_*$ in each Krylov subspace. For systems of linear size $L$, we first show that the critical disorder strength for a single-boson ($N_b=1$) sector tends to zero as $W_*\propto 1/\sqrt{L}\rightarrow 0$. Generically, we develop analytical arguments which imply $W_*$ remains finite as long as the boson number $N_b\lesssim L^\gamma/\log L$ ($1/2\le \gamma \le 1$) as system size $L\rightarrow\infty$. Since the area of the system scales as $L^2$, this implies that the boson number density $\rho_b\sim N_b/L^2$ has to tend to zero in the thermodynamic limit to have a finite $W_*$ for MBL. This is in contrast to the 1d quantum breakdown model with small fermion flavor number per site \cite{lian2023}, where MBL shows up generically in most charge sectors. This agrees with the general trend that MBL is more difficult to arise in higher dimensions. Nevertheless, the results of our model points into a new direction of searching for MBL with sub-dimensional number of quantum particles interacting with classical degrees of freedom, or more generically models of hybridization between quantum and classical systems.

The rest of the paper is organized as follows. The model and subsystem symmetries are given in \cref{sec1}. We discuss the analytical structure and numerical ED results for Hilbert space fragmentation and MBL in one-boson charge sectors in \cref{sec-one-boson}, and then present an analysis based on numerical ED results for multi-boson charge sectors in \cref{sec-N-boson}. The concluding remarks on future studies are given in \cref{sec-discuss}.

\section{The 2d quantum breakdown Model}\label{sec1}

\subsection{The model setup}

\begin{figure}[tbp]
\centering
\includegraphics[width=3.4in]{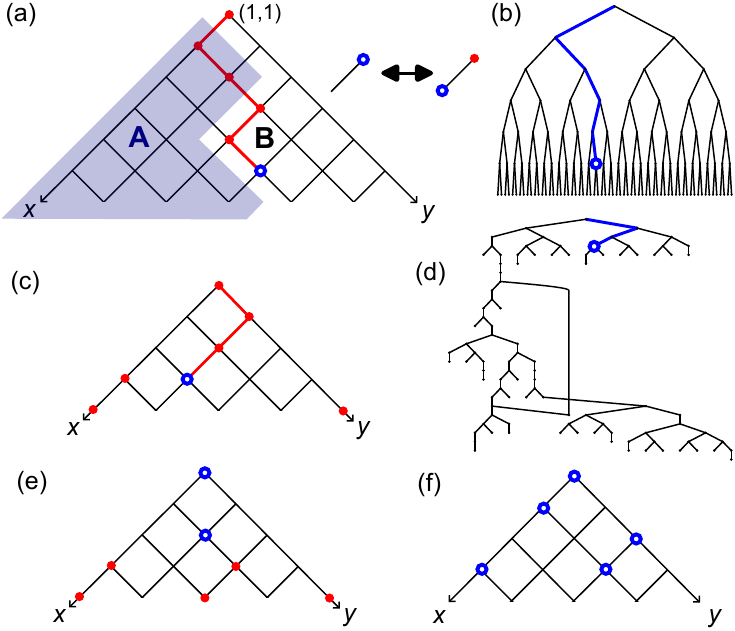}
\caption{(a) The real space lattice and subregions $A$, $B$. Blue circled, red dotted and empty sites denote $|2\rangle_{\zeta_\bR}$, $|1\rangle_{\zeta_\bR}$ and $|0\rangle_{\zeta_\bR}$, respectively. (a),(c),(e),(f) show Fock states (for small $L$) in four Krylov subspaces. (b) and (d) show the Fock space graphs of (a) and (c), respectively. In particular, panel (d) gives an example of graph with loops, which is due to the fact that the hardcore boson can either hop to a pre-existing spin up (red dot) site from below, or circumventing this site, when approaching a definite final state.}
\label{fig1-lattice}
\end{figure}

We define the 2d quantum breakdown model in a square lattice with sites $\bR=n_x \mathbf{e}_x+n_y\mathbf{e}_y=(n_x,n_y)$ as shown in \cref{fig1-lattice}(a), where $\mathbf{e}_x$ and $\mathbf{e}_y$ are $x$ and $y$ direction unit vectors, and $n_x,n_y\ge1$ are integers. Each site has a local spin $1/2$ with Pauli operators $\sigma_{x,\bR}$, $\sigma_{y,\bR}$, $\sigma_{z,\bR}$, and a spinless hardcore boson degree of freedom with creation/annihilation operators $b^\dag_\bR,b_\bR$ subject to the hardcore constraint $b^\dag_\bR b_\bR\le 1$. The Hamiltonian takes the form
\begin{equation}\label{eq-H}
\begin{split}
&H=\sum_\bR \Big[\epsilon_\bR \rho^\uparrow_{\bR} -\sum_{\nu=x,y}\left(\sigma_{+,\bR+\mathbf{e}_\nu}b^\dag_{\bR+\mathbf{e}_\nu}b_\bR \rho^\uparrow_{\bR} +h.c. \right)\Big],
\end{split}
\end{equation}
where $\rho^\uparrow_{\bR}=(1+\sigma_{z,\bR})/2$ is the projector into on-site spin $\uparrow$ state, and $\sigma_{\pm,\bR}=(\sigma_{x,\bR}\pm i\sigma_{y,\bR})/2$. 
The on-site energies $\epsilon_\bR$ are disordered and satisfy a uniform probability distribution in the interval
\begin{equation}
\epsilon_\bR\in \left[-W/2,W/2\right]\ ,
\end{equation}
where the disorder strength $W\ge0$. The second term in \cref{eq-H} is a homogeneous spatially asymmetric breakdown-type interaction \cite{lian2023,hu2024_boson,chen2024_breakdown,hu_gauge2024} between the spins and bosons, with interaction strength fixed to $1$. 

Such a model resembles the breakdown of a supersaturated vapor in a cloud chamber induced by incident particles. In this picture, each spin $\downarrow$ ($\uparrow$) represents an unperturbed (nucleated) vapor atom, and each hardcore boson plays the role of an incident particle roughly towards the $\mathbf{e}_x+\mathbf{e}_y$ direction, nucleating the vapor atoms in its path when moving in $+\mathbf{e}_x$ or $+\mathbf{e}_y$ direction. Such interactions where particles perturb other degrees of freedom in a spatially asymmetric way are generically termed as breakdown interactions \cite{lian2023,hu2024_boson,chen2024_breakdown,hu_gauge2024}. Our model is also similar to the quantum link model \cite{chandrasekharan1997,banerjee2012,kasper2017} which has spins on bonds and zero disorder instead.

Hereafter, we define a layer index 
\begin{equation}
l=n_x+n_y-1
\end{equation}
for each site $\bR=(n_x,n_y)$, and consider a lattice with $L$ layers restricted within a triangular region $n_x\ge 1$, $n_y\ge 1$ and $1\le l\le L$ as shown in \cref{fig1-lattice}(a). 
For convenience, we relabel each site $\bR$ by an integer 
\begin{equation}
\zeta_\bR=\frac{l^2+n_y-n_x+1}{2}\ , 
\end{equation}
which sorts all the sites from left to right layer by layer. Taking $L\rightarrow \infty$ gives the thermodynamic limit. 

We denote the unexcited state on site $\bR$ with spin $\downarrow$ and zero boson as $|0\rangle_{\zeta_\bR}$, and define three other on-site states $|1\rangle_{\zeta_\bR}=\sigma_{+,\bR}|0\rangle_{\zeta_\bR}$, $|2\rangle_{\zeta_\bR}=b^\dag_\bR\sigma_{+,\bR}|0\rangle_{\zeta_\bR}$ and $|3\rangle_{\zeta_\bR}=b^\dag_\bR|0\rangle_{\zeta_\bR}$. Particularly, state $|3\rangle_{\zeta_\bR}$ is decoupled from other on-site states in \cref{eq-H}, and will not be accessed if there are no such sites initially. Therefore, we shall simplify our model by constraining it into the Hilbert space of three on-site basis states $\left(|2\rangle_{\zeta_\bR},|1\rangle_{\zeta_\bR},|0\rangle_{\zeta_\bR}\right)$, denoted by blue-circled, red-dotted and empty sites in \cref{fig1-lattice}, respectively. More explicitly, the reduced Hamiltonian of this simplified model takes the form
\begin{equation}\label{eq-H3state}
\begin{split}
&H=\sum_\bR \Big[\epsilon_\bR \rho^\uparrow_{\bR}- \sum_{\nu=x,y}\left(A^\dag_{\bR+\mathbf{e}_\nu}B_{\bR} +h.c. \right) \Big] \ ,
\end{split}
\end{equation}
where the operators on site $\bR$ are defined in the basis  $\left(|2\rangle_{\zeta_\bR},|1\rangle_{\zeta_\bR},|0\rangle_{\zeta_\bR}\right)$ as
\begin{equation}
\begin{split}
&\rho^\uparrow_{\bR}=\left(
\begin{array}{ccc}
1&0&0\\
0&1&0\\
0&0&0\\
\end{array}
\right),
A_{\bR}^\dag=\left(
\begin{array}{ccc}
0&0&1\\
0&0&0\\
0&0&0\\
\end{array}
\right), B_{\bR}=\left(
\begin{array}{ccc}
0&0&0\\
1&0&0\\
0&0&0\\
\end{array}
\right)\ .
\end{split}
\end{equation}
Hereafter, we shall consider the simplified model \cref{eq-H3state}.

\subsection{Subsystem symmetry charges}

The simplified model of \cref{eq-H3state} has a set of subsystem symmetry charges, which can be derived as follows.

We assign the three states $\{|2\rangle_{\zeta_\bR},|1\rangle_{\zeta_\bR},|0\rangle_{\zeta_\bR}\}$ in a site $\bR$ in layer $l$ three different U(1) charges $\{q^l,q^l(1-q),0\}$, respectively, where $q$ is an arbitrary chosen number. Note that the interaction term $A^\dag_{\bR+\mathbf{e}_\nu}B_{\bR}+h.c.$ in \cref{eq-H3state} changes an on-site state $|2\rangle_\bR$ in layer $l$ with charge $q^l$ into a state $|1\rangle_{\zeta_\bR}$ in layer $l$ with charge $q^l(1-q)$ and a state $|2\rangle_{{\zeta_{\bR'}}}$ in layer $l+1$ with charge $q^{l+1}$. Since $q^l=q^l(1-q)+q^{l+1}$, the total U(1) charge is unchanged during this process for any $q$. Therefore, the total U(1) charge of the lattice
\begin{equation}\label{seq-Cq}
C(q)=\sum_{l=1}^L \left[ q^l N_{2,l}+ q^l(1-q)N_{1,l}\right] =\sum_{m=1}^{L+1} q^m Q_m
\end{equation}
is conserved, where we have defined $N_{j,l}=\sum_{\bR\in l}|j\rangle \langle j|_{\zeta_\bR}$ as the number of sites in layer $l$ which are in on-site state $|j\rangle_{\zeta_\bR}$ ($j=1,2$). In the rewritten form of power series of $q$, the coefficients $Q_m$ ($1\le m\le L+1$) in \cref{seq-Cq} can be derived as
\begin{equation}\label{seq-charge}
\begin{split}
&Q_1=N_{2,1}+N_{1,1}\ ,\\
&Q_m=N_{2,m}+N_{1,m}-N_{1,m-1}, \quad (2\le m\le L) \\
&Q_{L+1}=-N_{1,L}\ .
\end{split}
\end{equation}
Since $q$ is an arbitrary number, we conclude that each $Q_m$ in \cref{seq-charge} has to be a conserved charge, to make all $C(q)$ conserved. Note that each $Q_m$ is a \emph{subsystem symmetry conserved charge} involving only at most two layers $m$ and $m+1$. In particular, the boson number 
\begin{equation}
N_b=\sum_\bR b^\dag_\bR b_\bR =\sum_{l=1}^L N_{2,l} =\sum_{m=1}^{L+1} Q_m
\end{equation}
is a global conserved charge.

As we will show below, every charge sector fragments into extensive Krylov subspaces, each exhibiting MBL under disorders. Generically, a Krylov subspace is the Hilbert space spanned by states 
\begin{equation}
\{H^n|\psi\rangle,n\in\mathbb{Z},n\ge 0\}\ , 
\end{equation}
generated by acting Hamiltonian $H$ on a certain root state $|\psi\rangle$ (which is always chosen as a Fock state here).

\section{A simple one-boson charge sector}\label{sec-one-boson}

We first consider the Krylov subspaces of a simple charge sector with 
\begin{equation}\label{eq-simple-sector}
Q_m=\delta_{m,1}\ , 
\end{equation}
which has boson number $N_b=1$. For a $L$-layer lattice, we have $N_{1,L}=0$, $N_{2,m}=N_{1,m-1}-N_{1,m}$, and $N_{2,1}=1-N_{1,1}$. Consider occupation configurations with $N_{2,m}=\delta_{m,l}$ $(1\le l\le L)$, for which we would have $N_{1,m}=1$ if $m<l$, and $N_{1,m}=0$ if $m\ge l$. Therefore, in each layer $m\le l$, there is one site not in state $|0\rangle_{\zeta_\bR}$. Since there are $m$ sites in layer $m$, the position of this site has $m$ choices, and thus in total there are $\prod_{m=1}^l m=l!$ configurations with $N_{2,m}=\delta_{m,l}$. Thus, the total Hilbert space dimension of this charge sector is
\begin{equation}\label{eq-dim-1boson}
\text{dim}\mathcal{H}_{\{Q_m=\delta_{m,1}\}}=\sum_{l=1}^L l! >L!\sim L^L\ .
\end{equation}


\subsection{The largest Krylov subspace}

The largest Krylov subspace $\mathcal{K}_1$ in this sector has its root state given by 
\begin{equation}\label{eq-root-K1}
|\psi_{1}\rangle =|2\rangle_{1}\otimes\prod_{\zeta_\bR\neq 1}|0\rangle_{\zeta_\bR}\ , 
\end{equation}
which has a boson occupying site $(1,1)$ (label $\zeta_{(1,1)}=1$), and has state $|0\rangle_{\zeta_\bR}$ on all other sites. Since the boson excites the sites in its path to states $|1\rangle_\bR$ when moving in $+\mathbf{e}_x, +\mathbf{e}_y$ directions, and vice versa, every Fock state in this Krylov subspace has one-to-one correspondence with a path $\bP$ from site $(1,1)$ to the site the boson reaches, with all the sites along the path excited into states $|1\rangle_\bR$, as shown by the red line in \cref{fig1-lattice}(a). By moving in $+\mathbf{e}_x$ or $+\mathbf{e}_y$, each path reaching layer $l$ can extend to two different paths reaching layer $l+1$. Therefore, all the Hamiltonian $H$ connected Fock states, labeled by such paths, can be mapped to sites of a Cayley tree graph $\mathcal{G}_1$ of $L$ layers with coordination number $z=3$ as shown in \cref{fig1-lattice}(b). 
This thus gives a Krylov subspace $\mathcal{K}_{1}$ of dimension 
\begin{equation}\label{eq-dim-K1}
\text{dim}\mathcal{K}_{1}=\sum_{l=1}^{L} 2^{l-1}=2^L-1\ ,
\end{equation}
which is equal to the number of Cayley tree sites.
The Hamiltonian $H$ is closed when acting in this Krylov subspace, and maps to a single-particle tight-binding model in the Cayley tree graph $\mathcal{G}_1$ in \cref{fig1-lattice}(b): 
\begin{equation}\label{eq-HG}
\begin{split}
&H_{\mathcal{G}_1}=\sum_{\bP\in \mathcal{G}_1}V_\bP |\bP\rangle\langle {\bP}| -\sum_{\langle\bP\bP'\rangle\in \mathcal{G}_1}|\bP\rangle\langle {\bP'}|\ , \\
\end{split}
\end{equation}
where $\bP$ denotes a site of the graph $\mathcal{G}_1$ and corresponds to a real space path, $|\bP\rangle$ represents the Fock state mapping to the path, $\langle\bP\bP'\rangle$ represents connected pairs of sites in graph $\mathcal{G}_1$, and 
\begin{equation}\label{eq-VP1}
V_\bP=\sum_{\bR\in\bP}\epsilon_\bR 
\end{equation}
is the sum of the on-site energies $\epsilon_\bR$ of real space sites $\bR$ on path $\bP$.

\begin{figure}[tbp]
\centering
\includegraphics[width=3.4in]{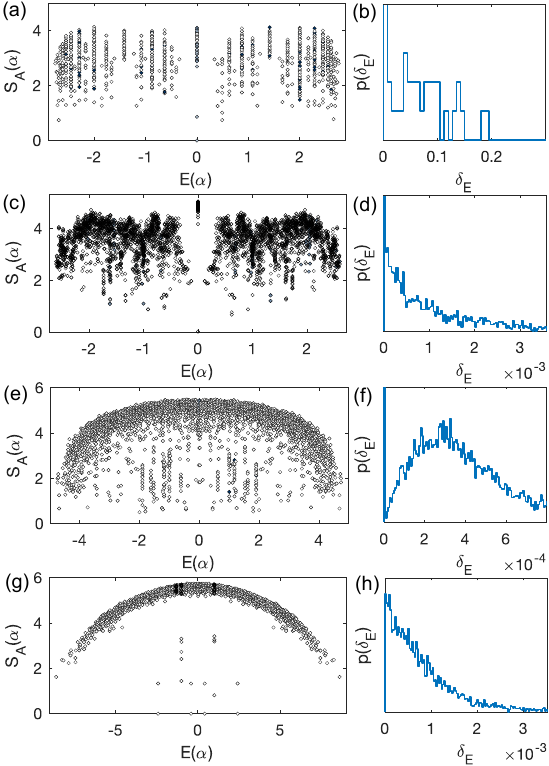}
\caption{ Numerical results at $W=0$ for $4$ Krylov subspaces illustrated in \cref{fig1-lattice}. Panels (a)-(b) corresponds to configuration of Krylov subspace in \cref{eq-root-K1} (\cref{fig1-lattice}(a)) with layer number $L=13$;  Panels (c)-(d) corresponds to configuration in \cref{eq-gen-1b} (\cref{fig1-lattice}(c)) with $L=7$; Panels (e)-(f) corresponds to configuration in the first example below \cref{eq-gen-nb} (\cref{fig1-lattice}(e)) with $L=7$; Panels (g)-(h) corresponds to configuration in the second example below \cref{eq-gen-nb} (\cref{fig1-lattice}(f)) with $L=5$. (a),(c),(e),(g) show their eigenstate entanglement entropy $S_A(\alpha)$ versus eigen-energy $E(\alpha)$, while (b),(d),(f),(h) show their LSS $p(\delta_E)$.}
\label{fig-W0}
\end{figure}

\subsubsection{Zero disorder: delocalized exact solution}

At zero disorder strength $W=0$, one has $V_\bP=0$, and the model \cref{eq-HG} in Cayley tree, or equivalently the Hamiltonian restricted within Krylov subspace $\mathcal{K}_1$, is known to be integrable \cite{ostilli2022}, and the many-body eigenstates are \emph{delocalized}. Moreover, in this case, the Krylov subspace $\mathcal{K}_1$ further fragments into smaller Krylov subspaces. Here we derive the eigenstates and eigen-energies.

Since each Fock state in the Cayley tree corresponds to a path $\bP$ of length no larger than $L-1$ in the real space, as explained below, for convenience we can relabel each Fock state $|\bP\rangle$ with a binary string of length $L$ of the form
\begin{equation}\label{eq-K1-binary-notation}
|n_1 n_2\cdots n_L\rangle \ ,
\end{equation}
where $n_j=0$ or $1$. For a path of $l-1$ steps which reaches layer $l$, the rightmost nonzero digit is defined to be $n_l=1$. For all $l'>l$, one has $n_{l'}=0$. For $l'<l$, we define $n_{l'}=1$ ($n_{l'}=0$) if the $l'$-th step of the path moves in the $\mathbf{e}_x$ ($\mathbf{e}_y$) direction. This gives a unique labeling of all the $2^L-1$ paths (Fock states) in Krylov subspace $\mathcal{K}_1$.

We now observe that, at $W=0$, the Krylov subspace $\mathcal{K}_1$ further fragments into smaller Krylov subspaces. These smaller Krylov subspaces are generated by root states orthogonal to each other. 
Except for the first root state in \cref{eq-root-K1} which in the new notation of \cref{eq-K1-binary-notation} takes the form
\begin{equation}\label{seq-root-W=0-1}
|\psi^{1}\rangle =|100\cdots 0\rangle\ ,
\end{equation}
any other generic root state is a state of superposition of two paths reaching layer $l_0\ge 2$ (path of $l_0-1$ steps):
\begin{equation}\label{seq-root-W=0}
\begin{split}
&|\psi^{l_0}_{n_1\cdots n_{l_0-2}}\rangle =\sum_{n_{l_0-1}=0}^1 \frac{(-1)^{n_{l_0-1}}}{\sqrt{2}}|n_1\cdots n_{l_0-1}10\cdots 0\rangle,
\end{split}
\end{equation}
where $n_1n_2\cdots n_{l_0-2}$ is a binary string consisting of $0$ and $1$. For $l_0\ge 2$, since the two Fock state components have opposite signs, one can show this root state in \cref{seq-root-W=0} cannot hop to any states of paths reaching layer $l<l_0$, thus $l_0$ is the smallest layer the state can hop to. But the root state in \cref{seq-root-W=0} can hop to $L-l_0$ other states $|\psi^{l_0,l}_{n_1n_2\cdots n_{l_0-2}}\rangle$ in larger layers $l>l_0$, which takes the following form:
\begin{equation}
\begin{split}
&|\psi^{1,l}\rangle=\frac{1}{2^{(l-1)/2}}\sum_{n_{l_0-1},\cdots,n_{l-1}=0}^1 |n_1 n_2\cdots n_{l-1}100\cdots 0\rangle, \\
&|\psi^{l_0,l}_{n_1n_2\cdots n_{l_0-2}}\rangle=\sum_{n_{l_0-1},\cdots,n_{l-1}=0}^1 \frac{(-1)^{n_{l_0-1}}}{2^{(l+1-l_0)/2}}\\
&\qquad\qquad  \times |n_1 n_2\cdots n_{l_0-2}n_{l_0-1}\cdots n_{l-1}100\cdots 0\rangle\ ,
\end{split}
\end{equation}
where $l\ge 1$ in the first line, and $l\ge l_0\ge 2$ in the second line. For a fixed $l_0$, the $L-l_0+1$ states $|\psi^{l_0,l}_{n_1n_2\cdots n_{l_0-2}}\rangle$ (with $l_0\le l\le L$) form a closed Krylov subspace, in which the zero disorder Hamiltonian $H$ takes the form of a one-dimensional open boundary tight-binding model with nearest hopping amplitude $-\sqrt{2}$. Therefore, the eigen-energies of this subspace is given by
\begin{equation}
E_{l_0,n}=-2\sqrt{2}\cos\left(\frac{n\pi}{L+2-l_0}\right)\ .
\end{equation}
These eigenstates are thus delocalized in the real space. The level degeneracy is given by the number of different root states in layer $l_0$ in \cref{seq-root-W=0}, which is the number of choices of the binary string $n_1n_2\cdots n_{l_0-2}$. This gives a level degeneracy $2^{l_0-2}+\frac{1}{2}\delta_{l_0,1}$ (note the fact that the degeneracy is $1$ when $l_0=1$.)

For the purpose of probing localization of a quantum state, we numerically calculate its entanglement entropy $S_A$ in the half subregion $A$ as shown in \cref{fig1-lattice}(a) by ED. \cref{fig-W0}(a) shows the ED numerical entanglement entropy $S_A(\alpha)$ of all the eigenstates $|\alpha\rangle$ at disorder $W=0$, where integer $\alpha$ sorts their energies $E(\alpha)$ from low to high. \cref{fig-W0}(b) shows the level spacing statistics (LSS) $p(\delta_E)$ of nearest neighbor energy level spacings 
\begin{equation}
\delta_E(\alpha)=E(\alpha+1)-E(\alpha)\ . 
\end{equation}
Both results show irregular patterns because of the integrability at $W=0$. The huge level degeneracy yields a sharp delta function peak at $\delta_E=0$.

\subsubsection{Nonzero disorder: MBL transition}\label{sec:1boson-Anderson}

To study the localization at nonzero disorder strength $W>0$, we first recall the standard Anderson localization model \cite{anderson1958}, which assumes random on-site potentials $V_\bP$ uniformly distributed in an interval $[-\frac{\Lambda}{2},\frac{\Lambda}{2}]$. In the infinite $z=3$ Bethe lattice, analytical and numerical studies \cite{abou1973,abou1974,mirlin1997,chalker1990,savitz2019,baroni2023} demonstrate the eigenstates become localized when 
\begin{equation}
\Lambda>\Lambda_*\approx 16\text{-}18\ .
\end{equation}
In the Cayley tree (finite patch of Bethe lattice), localization is delicate due to finite ratio between boundary and bulk sites in the thermodynamic limit. In the presence of boundary (i.e., last row of \cref{fig1-lattice}(b)), the above localization transition is difficult to identify. As shown in \cite{sade2003}, an effective way to get rid of the boundary effect is to randomly connect the boundary sites of a Cayley tree into coordination number $z=3$, with the hopping between all connected sites set to the same number ($-1$ in our model \cref{eq-HG}). With such a randomly connected boundary, an Anderson localization transition around $\Lambda_*$ can be observed \cite{sade2003} (see \cref{app-Cayley}). 


Our model \cref{eq-HG} with $W>0$ defines a \emph{modified} single-particle localization model with random potentials $V_\bP$ in \cref{eq-VP1}, where $\epsilon_\bR\in [-\frac{W}{2},\frac{W}{2}]$. Therefore, a particle traveling from site $\bP$ to another site $\bP'$ distance $d_{\bP,\bP'}$ away in the Cayley tree graph feels a random potential difference $|V_\bP-V_{\bP'}|\sim \sqrt{d_{\bP,\bP'}}W$. With the mean site distance $\langle d_{\bP,\bP'}\rangle \sim L$ in the Cayley tree, we estimate the effective on-site disorder strength $\Lambda\sim |V_\bP|_\text{rms}=\sqrt{\langle V_\bP^2\rangle}$ (root mean square of $V_\bP$) as
\begin{equation}\label{eq-lambda-1boson-cayley}
\Lambda\sim \sqrt{\langle d_{\bP,\bP'}\rangle}W\sim \sqrt{L}W\ ,
\end{equation}
which we have verified numerically (see \cref{app-scaling}). This suggests localization (MBL) in this Krylov subspace $\mathcal{K}_1$ to happen when 
\begin{equation}\label{eq-cayley-MBL}
W>W_*\sim \frac{\Lambda_*}{\sqrt{L}} \ ,
\end{equation}
with $\Lambda_*\approx 16\text{-}18$ defined earlier. Thus, $W_*\rightarrow 0$ in the thermodynamic limit $L\rightarrow\infty$.



\begin{figure}[tbp]
\centering
\includegraphics[width=3.4in]{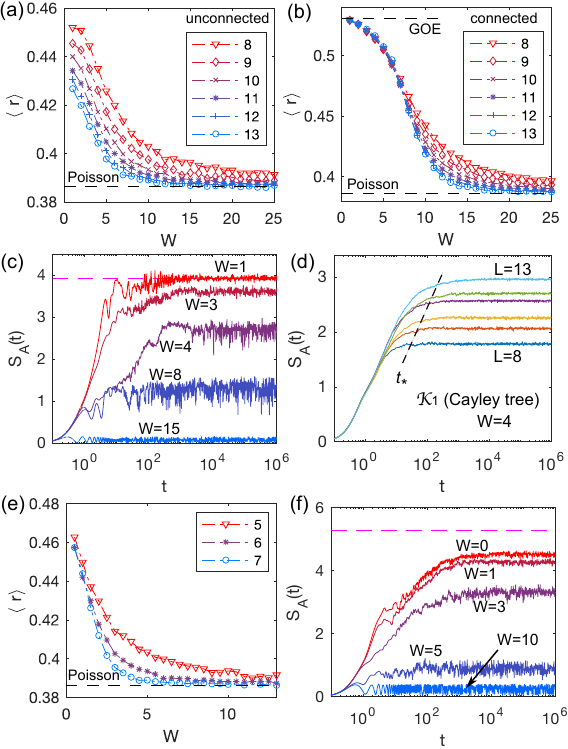}
\caption{For one-boson Krylov subspace $\mathcal{K}_1$ which maps to Cayley tree, the LSS ratio $\langle r\rangle$ with (a) unconnected and (b) connected boundary; and the time-evolved entanglement entropy $S_A(t)$ (unconnected boundary) from the root state for (c) different $W$ at $L=13$, and (d) different $L$ at $W=4$ ($100$-sample average). (e)-(f) show $\langle r\rangle$ and $S_A(t)$ (for $L=7$) of the one-boson Krylov subspace $\mathcal{K}_{1,\mathcal{S}_1}$ (illustrated in \cref{fig1-lattice}(c)-(d)). The legends in (a),(b),(e) label the layer number $L$.}
\label{fig-1p}
\end{figure}

Numerically, MBL would obey Poisson LSS $p(\delta_E)\propto e^{-\delta_E/\lambda_0}$ \cite{berry1977}, while the quantum chaotic thermalizing phase would obey Wigner-Dyson LSS (Gaussian orthogonal ensemble (GOE) for real Hamiltonian) $p(\delta_E)\propto\delta_E e^{-(\delta_E/\lambda_0)^2}$ \cite{bohigas1984,Wigner1967,Dyson1970}, with $\lambda_0$ some constant. We calculate the LSS ratio $\langle r\rangle$ defined as the mean value of \cite{oganesyan2007}
\begin{equation}
r(\alpha)=\frac{\min\{\delta_E(\alpha),\delta_E(\alpha+1)\}}{\max\{\delta_E(\alpha),\delta_E(\alpha+1)\}}\ ,
\end{equation}
which approaches $0.53$ for GOE and $0.39$ for Poisson \cite{atas2013}. 

\cref{fig-1p}(a) shows $\langle r\rangle$ (averaged over disorder samplings) with respect to $W$ in Krylov subspace $\mathcal{K}_{1}$, namely the Cayley tree with unconnected boundary sites, for different $L$ (up to $13$, given in legend), which stays below the GOE value for all $W$, making it difficult to determine the localization threshold. This is again due to the boundary effect of Cayley tree, similar to that shown in the conventional Anderson model in Cayley tree \cite{sade2003}. To eliminate the boundary effect, we follow \cite{sade2003} to modify the graph by randomly connecting the Cayley tree boundary sites into coordination number $z=3$, namely, each boundary site (the last row of \cref{fig1-lattice}(b)) is randomly connected with two other boundary sites (with all connected sites having the same hopping $-1$ in between as in \cref{eq-HG}).  \cref{fig-1p}(b) shows $\langle r\rangle$ calculated for $\mathcal{K}_{1}$ with randomly connected boundaries, which decreases from GOE to Poisson as $W$ increases. For $W>0$ ($W\gtrsim 5$) with unconnected (randomly $z=3$ connected) boundaries, the $\langle r\rangle$ curves monotonically decrease towards Poisson with increasing $L$. This agrees with our expectation $W_*\sim \frac{\Lambda_*}{\sqrt{L}}$. 

MBL in $\mathcal{K}_1$ at sufficiently large $W>W_*$ manifests more clearly in the entanglement entropy $S_A(t)$ versus time $t$ with initial state being the root state $|\psi_{1}\rangle$. As shown in \cref{fig-1p}(c) where $L=13$, when $W$ is small, such as $W=1$, the entanglement entropy $S_A(t)$ grows linearly in $t$ at small time and quickly satuates to dashed horizontal line. Hereafter, the dashed horizontal line is the Page value $S_\text{Page}$ calculated from the entanglement entropy of random states in the corresponding Krylov subspace (here $\mathcal{K}_1$), representing the entanglement entropy of thermalizing states. For $W$ sufficiently large $W\gtrsim 4$, there exist a saturation time $t_*$ such that
\begin{equation}\label{eq-SA-growth}
S_A(t)\approx\begin{cases}
& c_A \log t \ ,\qquad\qquad\quad   (t<t_*) \\
& \text{const.}<S_\text{Page}\ ,\qquad (t>t_*)
\end{cases}
\end{equation}
This behavior is clearly as expected for MBL. This agrees with our estimation in \cref{eq-cayley-MBL}, which yields $W_*\approx 3\text{-}4$ for $L=13$. In the MBL phase, for fixed $W$, the coefficient $c_A$ in \cref{eq-SA-growth} is roughly independent of $L$, and the saturation time $t_*\sim e^{L/\xi}$ with some $\xi>0$, as shown in the example of \cref{fig-1p}(d) which calculates the $L$-dependence of $S_A(t)$ at $W=4$ (each curve is calculated by averaging over $100$ random disorder samplings).


\subsection{All the other Krylov subspaces}

From \cref{eq-dim-1boson,eq-dim-K1}, we see that 
\begin{equation}
\frac{\text{dim}\mathcal{K}_{1}}{\text{dim}\mathcal{H}_{\{Q_m=\delta_{m,1}\}}}\sim (2/L)^L \ ,
\end{equation}
which indicates that the Krylov subspace $\mathcal{K}_{1}$ is an exponentially small subspace of the one-boson charge sector $\mathcal{H}_{\{Q_m=\delta_{m,1}\}}$. As we will show below, all the other Krylov subspaces have smaller dimensions $2^{L+1-l_K}-1$ ($l_K>1$), which have a Cayley tree structure similar to the largest Krylov subspace $\mathcal{K}_{1}$ and thus show similar MBL. 

Generically, we define the root state of a Krylov subspace in the single-boson charge sector \cref{eq-simple-sector} as the Fock state with the boson located in the smallest layer as possible in this Krylov subspace. Assume this boson is in layer $l_K$, and we denote the root state as $|\psi_{l_K}\rangle$. 

First, by \cref{seq-charge}, the charge sector in \cref{eq-simple-sector} requires that
\begin{equation}
\begin{split}
&N_{2,l}=\delta_{l,l_K}\ ,\quad N_{1,l}=\begin{cases}
&1\ ,\quad (l<l_K) \\
&0\ ,\quad (l\ge l_K)
\end{cases}\ .
\end{split}
\end{equation}
Namely, each layer $l<l_K$ contains only one site in in state $|1\rangle_{\zeta_\bR}$, while all the other sites except for the site of the boson (the site of state $|2\rangle_{\zeta_\bR}$) have to be in state $|0\rangle_{\zeta_\bR}$. 

Secondly, the site of boson (state $|2\rangle_{\zeta_\bR}$) in layer $l_K$ should not be the nearest neighbor of the site of state $|1\rangle_{\zeta_\bR}$ in layer $l_K-1$; otherwise the boson can hop back to the site of state $|1\rangle_{\zeta_\bR}$ in layer $l_K-1$, and $l_K$ would not be the smallest layer the boson can stay in, violating the assumption that $|\psi_{l_K}\rangle$ is a root state.


Now starting from a root state $|\psi_{l_K}\rangle$, when the boson moves to larger layers $l>l_K$ (which are initially all in state $|0\rangle_{\zeta_\bR}$), each such root state $|\psi_{l_K}\rangle$ can generate a Krylov subspace $\mathcal{K}_{l_K}$ of a Cayley tree of $L-l_K+1$ layers, in which each Fock state maps to a path from the root state boson site in layer $l_K$ to larger layers $l\ge l_K$, exactly analogous to $\mathcal{K}_{1}$. Thus, this Krylov subspace has a Hilbert space dimension
\begin{equation}
\begin{split}
&\text{dim}\mathcal{K}_{l_K}=\sum_{l=l_K}^L 2^{l-l_K}=2^{L+1-l_K}-1\ .
\end{split}
\end{equation}

The number of Krylov subspaces with this dimension is given by the number of distinct root states $|\psi_{l_K}\rangle$ satisfying the above requirement, that the site of state $|2\rangle_{\zeta_\bR}$ in layer $l_K$ should not be neighboring to the site of state $|1\rangle_{\zeta_\bR}$ in layer $l_K-1$. For $l_K>1$, the counting of root state choices of sites $|2\rangle_{\zeta_\bR}$ and $|1\rangle_{\zeta_\bR}$ is as follows: (1) in each layer $l\le l_K-2$, one is free to set state $|1\rangle_{\zeta_\bR}$ to any of the $l$ sites, yielding $l$ choices; (2) in layer $l=l_K-1$, one has $l_K-2$ choices of site of state $|1\rangle_{\zeta_\bR}$ if the state $|2\rangle_{\zeta_\bR}$ in layer $l_K$ is the leftmost/rightmost site, and $l_K-3$ choices otherwise, which yields in total $(l_K-3)(l_K-2)+2(l_K-2)$ choices of states in layers $l_K-1$ and $l_K$. This gives the total number of such Krylov subspaces with root state boson in layer $l_K$:
\begin{equation}
\begin{split}
D_K(l_K)&=(l_K-2)![(l_K-3)(l_K-2)+2(l_K-2)]\\
&=l_K!-2(l_K-1)!\ ,
\end{split}
\end{equation}
For $l_K=1$, one has only $D_K(1)=1$ Krylov subspace, which is the subspace $\mathcal{K}_1$ we defined in \cref{eq-dim-K1}. As one can verify, the total dimension of all these Krylov subspaces is equal to the dimension of this charge sector in \cref{eq-dim-1boson}:
\begin{equation}
\sum_{l_K=1}^L D_K(l_K) \text{dim}\mathcal{K}_{l_K}=\text{dim}\mathcal{H}_{\{Q_m=\delta_{m,1}\}}\ .
\end{equation}
Therefore, we see that the number 
of Krylov subspaces in this charge sector is exponentially large. Moreover, at nonzero disorders $W>0$, all these Krylov subspaces map to the modified Anderson localization problem on Cayley tree as we discussed in \cref{sec:1boson-Anderson} above, thus will show MBL in the $L\rightarrow\infty$ limit.

\section{Generic charge sectors}\label{sec-N-boson}

In a generic charge sector, the Hilbert space structure is usually much more complicated and can only be studied numerically by ED. We find that most charge sectors still have an exponentially large number of Krylov subspaces, and each Krylov subspaces show MBL behaviors at strong disorders $W$. We analyze the details in the below.

\subsection{Generic one-boson charge sectors}


We first consider generic one-boson charge sectors different from the one in \cref{sec-one-boson}. A generic one-boson charge sector will have $N_b=\sum_{m=1}^{L+1} Q_m=1$, while $Q_m$ can be positive or negative as defined in \cref{seq-charge}. As an example, we consider the charge sector containing the following state:
\begin{equation}\label{eq-gen-1b}
\begin{split}
&|\psi_{1,\mathcal{S}_1}\rangle=|2\rangle_{1}\otimes\prod_{\zeta_\bR\in \mathcal{S}_1}|1\rangle_{\zeta_\bR} \otimes\prod_{\zeta_\bR\notin \{1\}\cup\mathcal{S}_1}|0\rangle_{\zeta_\bR}\ ,\\
&\quad \mathcal{S}_1=\{7+4k\ |\ k\ge0\}\ ,
\end{split}
\end{equation}
which has a boson on site $1$, and a set of sites $\mathcal{S}_1$ in state $|1\rangle_{\zeta_\bR}$ (\cref{fig1-lattice}(c)). We allow the total number of layers $L$ to vary (which affects the size of the set $\mathcal{S}_1$). Numerically, the above state serves as the root state generating a Krylov subspace $\mathcal{K}_{1,\mathcal{S}_1}$, which is a subspace of the corresponding one-boson charge sector. 

The Hamiltonian in this Krylov subspace $\mathcal{K}_{1,\mathcal{S}_1}$ maps to a model similar to \cref{eq-HG} in a Fock space graph $\mathcal{G}_{1,\mathcal{S}_1}$ as shown in \cref{fig1-lattice}(d) (in which we set $L=5$), where the potential 
\begin{equation}\label{eq-VP-general}
V_\bP=\sum_\bR \epsilon_\bR\langle \rho^\uparrow_\bR\rangle_\bP 
\end{equation}
for graph site $\bP$, with $\langle \rho^\uparrow_\bR\rangle_\bP=1$ or $0$ if site $\bR$ has spin $\uparrow$ or $\downarrow$ as defined below \cref{eq-H}. The general definition of $V_\bP$ in \cref{eq-VP-general} reduces to \cref{eq-VP1} when restricted to the Krylov subspaces of \cref{eq-simple-sector}. The graph is more complicated than a Cayley tree and has loops (see also \cref{app-loop}), although the average bulk coordination number remains $z\approx 3$. Heuristically, the presence of loops is because, the presence of pre-existing spin up sites may allow the hardcore boson to hop to these sites from below, or circumventing these sites, when propagating towards a definite final state (see the example given in \cref{app-loop}). Numerically, we find $\mathcal{G}_{1,\mathcal{S}_1}$ has mean graph site distance $\langle d_{\bP,\bP'}\rangle\sim L^3$ (\cref{app-scaling}), longer than the Cayley tree case. One may think we could estimate the root mean square graph site potential $|V_\bP|_\text{rms}$ by $\langle d_{\bP,\bP'}\rangle$ similar to \cref{eq-lambda-1boson-cayley}. However, since the total number of real space sites scales as $L^2$, $|V_\bP|_\text{rms}$ cannot exceed that of the summation of potentials of all the real space sites. Therefore, we estimate the effective graph on-site disorder strength by
\begin{equation}\label{eq-Vp-1b}
\Lambda\sim |V_\bP|_\text{rms}\sim \min(\sqrt{\langle d_{\bP,\bP'}\rangle}W,\sqrt{L^2}W)\sim LW\ .
\end{equation}
we verified such a scaling of graph potential numerically, as shown in \cref{app-scaling}. Thus, this leads us to expect MBL to happen when 
\begin{equation}\label{eq-1b-MBL}
W>W_*\sim \frac{\Lambda_*}{L}\ , 
\end{equation}
again suggesting $W_*\rightarrow 0$ as $L\rightarrow \infty$. Comparing \cref{eq-1b-MBL} with \cref{eq-cayley-MBL}, we see that the Krylov subspace of generic one-boson charge sectors may enter MBL more easily as $L$ increases. A heuristic understanding for this difference is, in a generic one-boson charge sector, there are additional spin $\uparrow$ sites blocking the motion of the boson, which enhances the tendancy of MBL.


\subsubsection{Zero disorder}

At zero disorder $W=0$, the presence of random spin $\uparrow$ sites in a generic one-boson Krylov subspace play the role of disorders, so it is already intriguing to examine signatures of localization of the boson at $W=0$. \cref{fig-W0}(c) and (d) show the eigenstate entanglement entropy $S_A(\alpha)$ and LSS of Krylov subspace $\mathcal{K}_{1,\mathcal{S}_1}$ at $W=0$ for $L=7$, respectively. In particular, \cref{fig-W0}(d) readily shows Poisson LSS, showing signature of integrability and the tendency of MBL. This is in contrast to the non-interacting particle in irregular shaped box (e.g., stadium billiard \cite{heller1984}) which shows Wigner-Dyson LSS and single-particle quantum chaos. Indeed, although here we only have one boson in the space, it is interacting with many spins (which are non-dynamical by themselves), making the LSS drastically different.

In addition, there are many intriguing sets of degenerate levels, as can be seen in \cref{fig-W0}(c) and indicated by the delta function peak at $\delta_E=0$ \cref{fig-W0}(d). These degenerate levels reveal some finer structure of the eigenstates at $W=0$, and possible analytical ways for solving the model, which is to be studied in the future.

\subsubsection{Nonzero disorder}

At nonzero disorder $W>0$, \cref{fig-1p}(e) shows that the LSS ratio $\langle r\rangle$ monotonically approaches Poisson as $L$ increases. For all $W\ge 0$, as shown by \cref{fig-1p}(f) (where $L=7$), the entanglement entropy time evolved from the root state $|\psi_{1,\mathcal{S}_1}\rangle$ shows $S_A(t)\propto \log t$ when $t<t_*$ and saturates below the Page value (dashed line) when $t>t_*$, similar to \cref{eq-SA-growth}, with $t_*\sim e^{L/\xi}$ (see \cref{app-EE}). This strongly supports the expected MBL for $W>W_*\rightarrow 0$ in the thermodynamic limit $L\rightarrow 0$, in agreement with our estimation in \cref{eq-1b-MBL}.

Heuristically, our results suggest that for Krylov subspaces with a single-boson, the presence of random spins in the space play the role of certain disorders even at $W=0$, making the system easier to enter the MBL phase (showing MBL features readily at small $W$). As we will see later, the multi-boson Krylov subspaces would be more difficult to localize.


\subsection{Multi-boson charge sectors} 

We now study Krylov subspaces in multi-boson charge sectors with $N_b>1$. To do this, we consider charge sectors containing the following multi-boson root states
\begin{equation}\label{eq-gen-nb}
|\psi_{\mathcal{S}_2,\mathcal{S}_1}\rangle =\prod_{\zeta_\bR\in \mathcal{S}_2}|2\rangle_{\zeta_\bR}\otimes\prod_{\zeta_\bR\in \mathcal{S}_1}|1\rangle_{\zeta_\bR} \otimes\prod_{\zeta_\bR\notin \mathcal{S}_2\cup\mathcal{S}_1}|0\rangle_{\zeta_\bR}\ , 
\end{equation}
where $\mathcal{S}_2$ and $\mathcal{S}_1$ are the sets of sites in $|2\rangle_{\zeta_\bR}$ and $|1\rangle_{\zeta_\bR}$ states, respectively. We then consider the Krylov subspaces $\mathcal{K}_{\mathcal{S}_2,\mathcal{S}_1}$ generated by the above multi-boson root states. We present the results for the following two representative examples of multi-boson Krylov subspaces: 

(1) the two-boson Krylov subspace generated by root state with 
\begin{equation}\label{eq-Kr-2b}
\mathcal{S}_2=\{1,5\}\ ,\qquad \mathcal{S}_1=\{7+2k\ |\ k\ge0\}\ ,
\end{equation}
as shown in \cref{fig1-lattice}(e); 

(2) the five-boson Krylov subspace generated by root state with 
\begin{equation}\label{eq-Kr-5b}
\mathcal{S}_2=\{1,2,6,7,9\}\ ,\qquad \mathcal{S}_1=\emptyset\ ,
\end{equation}
as shown in \cref{fig1-lattice}(f), where $\emptyset$ stands for the empty set. 

In both examples, we allow the layer number $L$ to vary (which affects the size of the set $\mathcal{S}_1$). Our numerical computation is limited up to $L=7$ ($L=5$) layers for the first (second) example above.

\subsubsection{Zero disorder: integrable, or chaotic with quantum scars}

At zero disorder $W=0$, we find a multi-boson Krylov subspace can generically show either Wigner-Dyson GOE (chaotic) or Poisson LSS, and quite often retains certain (potentially solvable) analytical structures. \cref{fig-W0}(e)-(h) shows the eigenstate entanglement entropy $S_A(\alpha)$ and LSS in the two-boson Krylov subspace in \cref{eq-Kr-2b} with $L=7$ and the five-boson Krylov subspace in \cref{eq-Kr-5b} with $L=5$, respectively. The two-boson Krylov subspace shows Wigner-Dyson GOE LSS (\cref{fig-W0}(f)), except for a delta function peak at zero implying the existence of degenerate states, which implies the Krylov subspace is quantum chaotic. In contrast, the five-boson Krylov subspace shows a Poisson LSS (\cref{fig-W0}(h)), which is potentially integrable. Intriguingly, in both Krylov subspaces, a subset of eigenstates show significantly lower entanglement entropy than the mojarity eigenstates. Moreover, among this subset of states, a significant fraction of states have degeneracies, as can be seen in \cref{fig-W0}(e),(g) and from the delta function peak at $\delta_E=0$ in \cref{fig-W0}(f). This implies certain solvable structures guaranteeing the level degeneracies.

A particularly interesting case is the example of two-boson Krylov subspace in \cref{eq-Kr-2b}, which is a chaotic Krylov subspace (indicated by Wigner-Dyson LSS) showing many-body \emph{quantum scar} states. As the eigenstate entanglement entropy $S_A(\alpha)$ in \cref{fig-W0}(e) shows, among the majority of eigenstates which show chaotic LSS and have volume law entanglement entropy, the subset of low-entanglement entropy eigenstates (many of which are degenerate) fulfill the definition of quantum scar states \cite{bernien2017,moudgalya2018,schecter2018,turner2018a,choi2019scar,ho2019,bull2019,lin2019,khemani2019,scherg2021}.

\begin{figure}[tbp]
\centering
\includegraphics[width=3.4in]{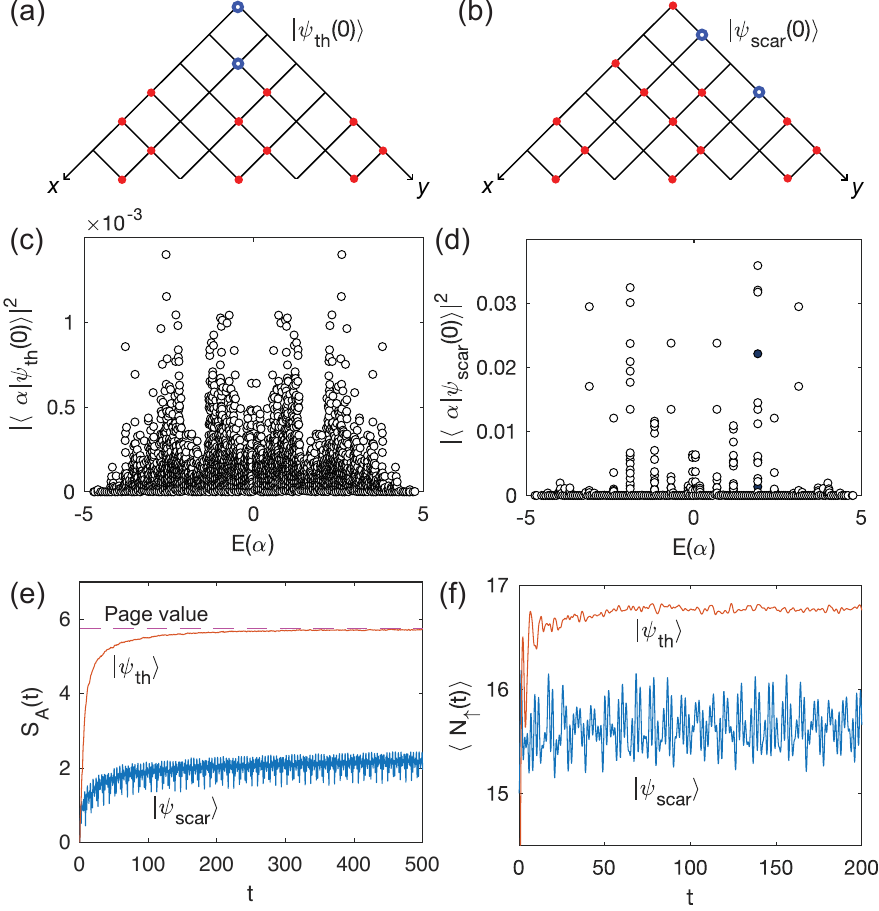}
\caption{The dynamics of thermalizing and non-thermalizing scar states in the two-boson Krylov subspace in \cref{eq-Kr-2b} at $W=0$, with the layer number $L=7$. The boson and spin configurations of the thermalizing and non-thermalizing initial Fock states $|\psi_\text{th}(0)\rangle$ and $|\psi_\text{scar}(0)\rangle$ are shown in (a) and (b), with blue hollow and red solid dots representing on-site state $|2\rangle_{\zeta_\bR}$ or $|1\rangle_{\zeta_\bR}$. (c) and (d) shows their wavefunction overlaps with the eigenstates $|\alpha\rangle$ of the Krylov subspace. (e) Their entanglement entropy $S_A(t)$ as a function of time $t$. The horizontal dashed line labels the Page value of the Krylov subspace. (f) Their total spin up site number $\langle N_\uparrow(t)\rangle$ as a function of time $t$. The scar initial state $|\psi_\text{scar}\rangle$ shows evident non-thermalizing behaviors.}
\label{figscar}
\end{figure}

These quantum scar states give intriguing non-thermalizing quantum dynamics for certain initial states. To see this, we set the layer number $L=7$, and consider two initial Fock states $|\psi_\text{th}(0)\rangle$ and $|\psi_\text{scar}(0)\rangle$ in this Krylov subspace of \cref{eq-Kr-2b} showing thermalizing dynamics and non-thermalizing scar dynamics, respectively. The thermalizing initial state is simply the root state $|\psi_\text{th}(0)\rangle=|\psi_{\mathcal{S}_{2},\mathcal{S}_{1}}\rangle$ given by \cref{eq-gen-nb}, with the two sets $\mathcal{S}_{2}$ and $\mathcal{S}_{1}$ given by \cref{eq-Kr-2b}. The non-thermalizing initial state $|\psi_\text{scar}(0)\rangle=|\psi_{\mathcal{S}_{2,\text{scar}},\mathcal{S}_{1,\text{scar}}}\rangle$ is a Fock state defined similar to \cref{eq-gen-nb}, with the two sets given by $\mathcal{S}_{2,\text{scar}}=\{3,10\}$ and $\mathcal{S}_{1,\text{scar}}=\{1,4,8,9\}\cup \{11+2k\ |\ k\ge0\}$. The boson and spin configurations of these two Fock states are given in \cref{figscar}(a) and (b), respectively. \cref{figscar}(c) and (d) show their overlaps $|\langle \alpha|\psi_\text{th}(0)\rangle|^2$ and $|\langle \alpha|\psi_\text{scar}(0)\rangle|^2$ with the energy eigenstates $|\alpha\rangle$ versus eigen-energies $E(\alpha)$, respectively, from which we see that $|\langle \alpha|\psi_\text{th}(0)\rangle|^2$ has large overlaps with most eigenstates, while $|\psi_\text{scar}(0)\rangle$ only dominantly overlaps with the low-entanglement entropy quantum scar eigenstates. 

We then calculate the dynamical time evolution of these two states. \cref{figscar}(e) shows that the entanglement entropy $S_A(t)$ of state $|\psi_\text{th}(t)\rangle$ quickly saturates the Page value of the Krylov subspace (dashed horizontal line), while $S_A(t)$ of state $|\psi_\text{scar}(0)\rangle$ saturates much lower than the Page value and shows persistent oscillations. We further examine the expectation value $\langle N_\uparrow(t)\rangle$ of the total number of spin $\uparrow$ sites (sites in state $|2\rangle_{\zeta_\bR}$ or $|1\rangle_{\zeta_\bR}$) defined as
\begin{equation}
N_\uparrow=\sum_\bR \rho^\uparrow_\bR\ ,
\end{equation}
with $\rho^\uparrow_\bR$ defined below \cref{eq-VP-general}. \cref{figscar}(f) shows $\langle N_\uparrow(t)\rangle$ as a function of time $t$ for both states. It is clear that the state $|\psi_\text{th}(t)\rangle$ has quickly thermalizing $\langle N_\uparrow(t)\rangle$ approaching an equilibriated value, while the state $|\psi_\text{scar}(0)\rangle$ shows non-thermalizing scar dynamics with persistent oscillations.

We note that such chaotic multi-boson Krylov subspaces with quantum scar states are not rare in our 2d quantum breakdown model at zero disorder $W=0$. Intriguingly, compared to the quantum scar states in the 1d quantum breakdown models \cite{lian2023,hu2024_boson,chen2024_breakdown,hu_gauge2024} which can be derived relatively analytically, the energy spectrum and degeneracies of the quantum scar states in the 2d quantum breakdown model here show much more irregular patterns. Also, unlike many previous quantum scar models such as the PXP model \cite{turner2018a,hu_gauge2024} which have energetically (almost) equally spaced scar states, the quantum scar energy levels in our model here are not equally spaced in energies. This is reflected in the scar dynamics as shown in \cref{figscar}(f), in which the persistent oscillations are not exactly periodic, signaling unequal energy spacings among the scar eigenstates. It would be interesting to further investigate the mechanism of this type of irregular quantum scar states in our 2d quantum breakdown model, and to explore other models showing similar quantum scar states.


\subsubsection{Nonzero disorder: generic MBL transition}

\begin{figure}[tbp]
\centering
\includegraphics[width=3.4in]{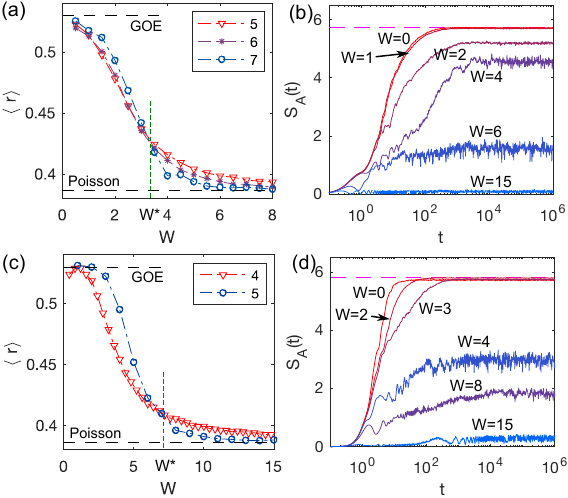}
\caption{The LSS ratio $\langle r\rangle$, and time-evolved entanglement entropy $S_A(t)$ starting from the root state, for: (a)-(b) the two-boson Krylov subspace illustrated in \cref{fig1-lattice}(e), and (c)-(d) the five-boson Krylov subspace illustrated in \cref{fig1-lattice}(f). The legends in (a),(c) label the layer number $L$. (b) is calculated for $L=7$, and (c) is calculated for $L=5$.}
\label{fig-np}
\end{figure}

At nonzero disorder $W>0$, in both the two-boson (\cref{eq-Kr-2b}) and five-boson (\cref{eq-Kr-5b}) Krylov subspaces, the LSS ratio $\langle r\rangle$ varies from GOE to Poisson as $W$ increases (\cref{fig-np}(a) and (c)), and the $\langle r\rangle$ curves for different $L$ (up the largest calculable $L$) intersect at certain $W=W_*$. The entanglement entropy time evolution from the root state shows $S_A(t)\propto t$ ($S_A(t)\propto \log t$) at early $t$ and saturation at (below) the Page value at late $t$ for small (large) $W$ (\cref{fig-np}(b) and (d)), similar to \cref{eq-SA-growth}. These features implies a transition from thermalization ($W<W_*$) to MBL ($W>W_*$), which we analyze below.

A $N_b$-boson Krylov subspace generically maps to a complicated Fock space graph $\mathcal{G}_{\mathcal{S}_2,\mathcal{S}_1}$ with numerous loops analogous to that in \cref{fig1-lattice}(d). The on-site potential $V_\bP$ in the Fock space graphs are still given by the generic formula in \cref{eq-VP-general}. The average coordination number can be estimated as
\begin{equation}
z\approx 2N_b+1 \ . 
\end{equation}
This is because each boson can approximately move to the next layer in two directions ($+\mathbf{e}_x$ and $+\mathbf{e}_y$) unless being blocked by a spin $\uparrow$, so the coordination number is estimated to be $z\approx 2N_b+1$ (the $1$ comes from the previous state generating this state). 

The presence of numerous complicated loops make it difficult to estimate the mean graph distance analytically. Instead, numerically, we find the mean graph distance in such Krylov subspaces scales in power law of system size $L$ as $\langle d_{\bP,\bP'} \rangle\sim L^{\eta}$ in $L$, with $\eta\approx 3$ in both the two-boson and five-boson Krylov subspaces here. This upper bounds the effective graph on-site disorder strength $\Lambda\sim |V_\bP|_\text{rms} \lesssim \min(\sqrt{\langle d_{\bP,\bP'}\rangle}W,\sqrt{L^2}W)$ as \cref{eq-Vp-1b} suggests. Numerically, we generically find the graph site potential scales as
\begin{equation}
|V_\bP|_\text{rms}\sim L^{\gamma}W \ , 
\end{equation}
with $\frac{1}{2}\le\gamma\le \min (\frac{\eta}{2},1)$. For the above two-boson and five-boson Krylov subspaces, $\gamma\approx1$ and $\frac{1}{2}$, respectively (see \cref{app-scaling}). Thus, we expect MBL to happen when $W>W_*\sim \frac{\Lambda_*(z)}{L^\gamma}$, where $\Lambda_*(z)$ is the standard Anderson transition point in graphs with coordination number $z$, which scales as $\Lambda_*(z)\propto z\log z\sim N_b\log N_b$ when $N_b$ is large \cite{abou1973,mirlin1997}, so we expect the MBL to happen when
\begin{equation}
W>W_*\sim \frac{N_b\log N_b}{L^\gamma}\ ,
\end{equation}
with $\frac{1}{2}\le\gamma\le 1$. Particularly, if the number of bosons 
\begin{equation}
N_b\lesssim \frac{L^\gamma}{\log L} \ ,
\end{equation}
namely, if the boson concentration is dilute enough, $W_*$ would be finite or zero when taking the thermodynamic limit $L\rightarrow \infty$.



\section{Discussion}\label{sec-discuss}

We have shown that our 2d quantum breakdown model strongly fragments into exponentially many Krylov subspace. At zero disorder strength $W=0$, we find each Krylov subspace is either potentially integrable (with Poisson LSS), or quantum chaotic (with Wigner-Dyson GOE LSS) with a set of many-body quantum scar states. An integrable example is the one-boson Krylov subspace $\mathcal{K}_1$ which we showed maps to a Cayley tree Fock space graph. In the chaotic Krylov subspaces, very interestingly, the quantum scar states exhibit irregular energy spacing and degeneracy patterns. This is in contrast to many previously studied quantum scar models which have (almost) equally spaced scar energy levels. Accordingly, the non-thermalizing scar quantum dynamics exhibit persistent oscillations not at a definite frequency. Further studies are desired to understand the mechanics of such generic quantum scar states.

At nonzero disorders, each Krylov subspace exhibits MBL beyond certain disorder strength $W_*$, with $W_*$ finite in the thermodynamic limit if $N_b\lesssim L^\gamma/\log L$ ($\frac{1}{2}\le \gamma\le 1$), where $N_b$ is the boson number and $L$ is the system size. In the simplest one-boson charge sector, we give an analytical understanding of the Hilbert space fragmentation into Krylov subspaces, each of which resembles a modified single-particle Anderson localization problem in the Cayley tree when viewed in the Fock space, which always localizes as $L\rightarrow\infty$. In generic charge sectors, each Krylov subspace also maps to the Anderson localization problem in a graph of the Fock space, in which the disorder strength threshold for MBL can be estimated in a controllable manner. 

Generically, the boson number density will scales as $\rho_b\sim N_b/L^2$, thus, to have the MBL critical strength $W_*$ finite, one requires $\rho_b\lesssim L^{\gamma-2}/\log L$ ($\frac{1}{2}\le \gamma\le 1$), which tends to zero in the thermodynamic limit $L\rightarrow \infty$. In contrast, in the 1d quantum breakdown model with small fermion flavor number studied in \cite{lian2023}, MBL is much more robust irrespective to the charge of the Krylov subspace. In general, it is expected that MBL in higher dimensions are more difficult to achieve. However, our results imply that a sub-dimensional number of bosons may have MBL in the thermodynamic limit. Heuristically, we can understand this MBL as emerging from quantum particles interacting with a swampland of spin configurations which would be non-dynamical (i.e., classical) without the particles. This may allow generalization into a class of MBL models, and provide understandings of generic preconditions for achieving higher dimensional MBL. This could also motivate the study of the stability of a lower dimensional MBL quantum system in a higher dimensional environment with classical degrees of freedom.

Moreover, it will be intriguing to explore potential realizations of such models in experiments such as Rydberg atoms, and the effects from perturbations respecting or breaking the subsystem symmetries. Lastly, as a simple quantum modeling of measurement devices like the cloud chamber, our model may be coupled to generic quantum systems, the study of which may provide a deeper understanding of quantum measurement therein.




\begin{acknowledgments}
\emph{Acknowledgments}. We thank David A. Huse and Yumin Hu for helpful discussions. This work is supported by the Alfred P. Sloan Foundation, the National Science Foundation through Princeton University’s Materials Research Science and Engineering Center DMR-2011750, and the National Science Foundation under award DMR-2141966. Additional support is provided by the Gordon and Betty Moore Foundation through Grant GBMF8685 towards the Princeton theory program.
\end{acknowledgments}











\appendix



\section{Comparing the Modified and the Standard Anderson Localization problem on Cayley tree}\label{app-Cayley}

\begin{figure*}[htbp]
\centering
\includegraphics[width=6.8in]{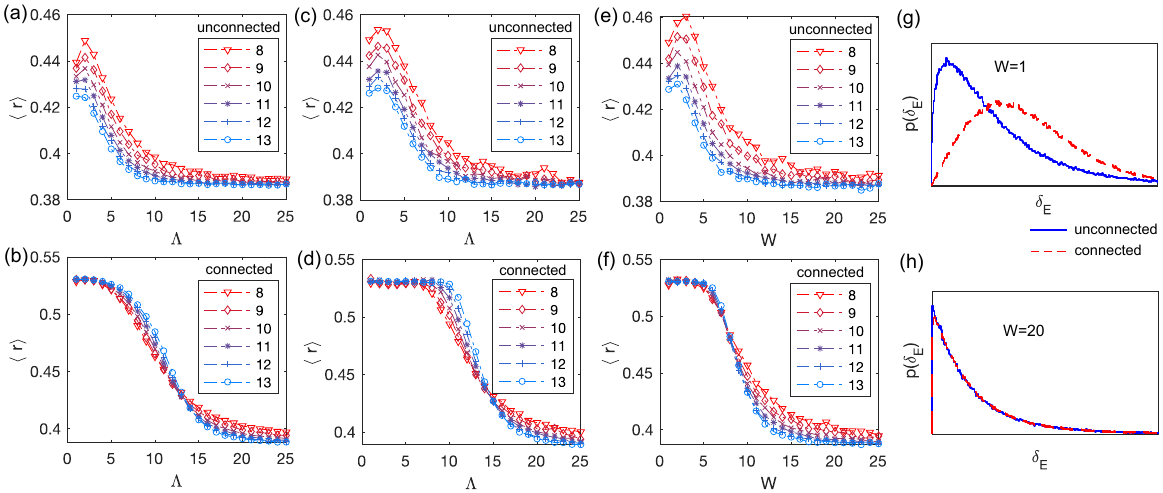}
\caption{(a)-(b) The full spectrum $\langle r\rangle$ and (c)-(d) the central spectrum $\langle r\rangle$ of the standard Anderson localization model in the Cayley tree, with unconnected or connected boundary. (e)-(f) The central spectrum $\langle r\rangle$ of our model in Krylov subspace $\mathcal{K}_1$ with unconnected or connected boundary. (g)-(h) The full spectrum LSS probability distribution of our model with unconnected or connected boundary (averaged over many disorder samplings).}
\label{figS-Anderson}
\end{figure*}

In the \cref{fig-1p}(a)-(b), we have shown the full spectrum LSS ratio $\langle r\rangle$ of our model in Krylov subspace $\mathcal{K}_1$, which maps to a modified Anderson localization problem in the Cayley tree in Fock space as shwon in \cref{eq-HG}. As $L$ increases (calculated up to $13$), the $\langle r\rangle$ curves with respect to disorder strength $W$ of different $L$ for unconnected boundary (which is our model) decrease towards Poisson, while those for randomly connected boundary (into coordination number $z=3$, which is modified on top of our model) decrease towards Poisson for $W\gtrsim 5$ and remain almost unchanged for $W\lesssim 5$. \cref{figS-Anderson}(g)-(h) show the LSS probability function $p(\delta_E)$ (averaged over many samplings) for unconnected (blue solid line) and connected (red dashed line) boundary, at $W=1$ and $W=20$, respectively. For unconnected boundary, the LSS deviates from GOE Wigner-Dyson function significantly at small $W$.

In comparison, for the standard Anderson localization problem in the Cayley tree, which has on-site potential in \cref{eq-HG} uniformly distributed in
\begin{equation}
V_\bP\in [-\Lambda/2,\Lambda/2]\ ,
\end{equation}
the LSS ratio $\langle r\rangle$ (averaged over many samplings) of the full spectrum with unconnected boundary is shown in \cref{figS-Anderson}(a), and that with boundary randomly connected into coordination number $z=3$ is shown in \cref{figS-Anderson}(b). As $L$ increases (calculated up to $13$), the unconnected boundary $\langle r\rangle$ curves with respect to $\Lambda$ also monotonically decreases, while the connected boundary $\langle r\rangle$ curves for different $L$ intersects around $\Lambda_*\approx 13$, signaling an Anderson transition (close to the Bethe lattice value $\Lambda_*\approx 16\text{-} 18$). This is similar to the observation in Ref. \cite{sade2003}.

The center of the energy spectrum is generically more difficult to localize. Therefore, we also calculate the central spectrum LSS ratio $\langle r\rangle$ of the energy levels within an energy window of width $0.1(E_\text{max}-E_\text{min})$ at the center of the spectrum, where $E_\text{max}$ and $E_\text{min}$ are the maximal and minimal energy of the spectrum, respectively. \cref{figS-Anderson}(c)-(d) shows the central spectrum $\langle r\rangle$ of the standard Anderson problem with unconnected and connected boundary, respectively. The connected boundary figure shows a transition point slightly higher, around $\Lambda_*\approx 14$. In comparison, \cref{figS-Anderson}(e)-(f) show the central spectrum LSS ratio of our model in Krylov subspace $\mathcal{K}_1$ (modified Anderson model in Cayley tree, \cref{eq-HG}) with unconnected and connected boundary, respectively. The connected boundary result \cref{figS-Anderson}(f) shows a transition around $W_*\approx 6\text{-} 7$. As argued in \cref{eq-cayley-MBL}, we expect $W_*\sim \frac{\Lambda_*}{\sqrt{L}}$, which is serves as a good estimation.

\ \\

\section{Example of loops in the Fock space graph}\label{app-loop}

\begin{figure*}[htbp]
\centering
\includegraphics[width=5.4in]{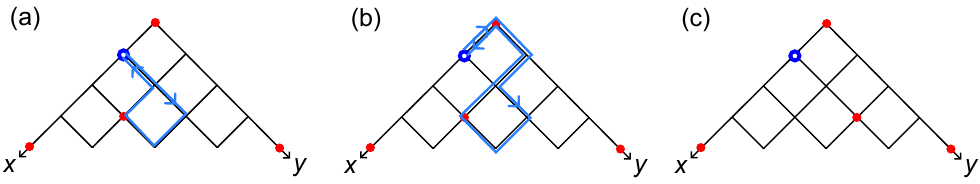}
\caption{Illustration of a loop in the 1-boson Krylov subspace $\mathcal{K}_{1,\mathcal{S}_1}$. Starting from the same Fock state, (a) and (b) show two different paths which end up at the same Fock state in (c).}
\label{figS-loop}
\end{figure*}

For the 1-boson Krylov subspace $\mathcal{K}_{1,\mathcal{S}_1}$ we studied, illustrated in \cref{fig1-lattice}(c), which is generated by root state $|\psi_{1,\mathcal{S}_1}\rangle=|2\rangle_{1}\otimes\prod_{\zeta_\bR\in \mathcal{S}_1}|1\rangle_{\zeta_\bR} \otimes\prod_{\zeta_\bR\notin \{1\}\cup\mathcal{S}_1}|0\rangle_{\zeta_\bR}$, with a set of sites $\mathcal{S}_1=\{7+4k\ |\ k\ge0\}$ in state $|1\rangle_{\zeta_\bR}$, its Fock space graph contain loops as shown in \cref{fig1-lattice}(d). As an example, for layer number $L=5$, \cref{figS-loop} shows how a loop in this Fock space graph looks like in the real space: (a) and (b) show two different paths of the boson from the same initial state, which end up at the same Fock state in (c).

More generically, Krylov subspaces with root states with multiple bosons and/or spin up sites can have many loops in the Fock space graph, due to the complicated interactions between bosons and the spins.

\section{System size scaling of Fock space graph distances and graph on-site potentials}\label{app-scaling}

\begin{figure*}[htbp]
\centering
\includegraphics[width=6.8in]{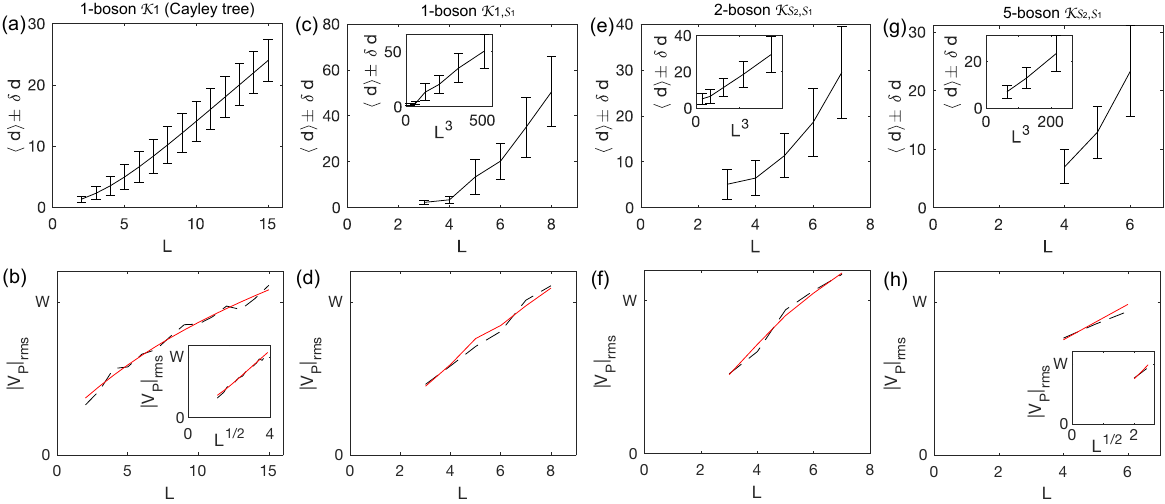}
\caption{For the four Krylov subspaces we considered in the main text (listed in \cref{seq-4Kry}, see also \cref{fig1-lattice}), (a),(c),(e),(g) show the numerical mean Fock-space graph distance $\langle d\rangle$ between two graph sites, with standard deviation $\delta d$, with respect to system size (number of layers) $L$. (b),(d),(f),(h) show the corresponding (the same column) root mean square value $|V_\bP|_\text{rms}=\sqrt{\langle V_\bP^2 \rangle }$ of graph on-site potential $V_\bP$ with respect to $L$.}
\label{figS-Lscaling}
\end{figure*}

For the  Fock space graphs of the four Krylov subspaces we considered in the main text (illustrated in \cref{fig1-lattice}):
\begin{widetext}
\begin{equation}\label{seq-4Kry}
\begin{split}
&\text{(1) one-boson Krylov (Cayley tree)  }\mathcal{K}_1\ , \\
&\text{(2) one boson Krylov}\qquad\qquad\qquad \mathcal{K}_{1,\mathcal{S}_1}\ ,\qquad \mathcal{S}_1=\{7+4k\ |\ k\ge0\} \\
&\text{(3) two-boson Krylov}\qquad\qquad\qquad  \mathcal{K}_{\mathcal{S}_2,\mathcal{S}_1}\ ,\qquad \mathcal{S}_2=\{1,5\}\ ,\quad \mathcal{S}_1=\{7+2k\ |\ k\ge0\} \\
&\text{(4) five-boson Krylov}\qquad\qquad\qquad  \mathcal{K}_{\mathcal{S}_2,\mathcal{S}_1}\ ,\qquad \mathcal{S}_2=\{1,2,6,7,9\}\ ,\quad \mathcal{S}_1=\emptyset \ ,
\end{split}
\end{equation}
\end{widetext}
we numerically calculate the mean graph distance $\langle d\rangle=\langle d_{\bP,\bP'}\rangle$ between two arbitrary graph sites $\bP$ and $\bP'$ and its standard deviation $\delta d$. The resulting $\langle d\rangle \pm \delta d$ for different system sizes $L$ are shown in \cref{figS-Lscaling}(a),(c),(e),(g), for the four Krylov subspaces in \cref{seq-4Kry}, respectively. In particular, we find $\langle d\rangle\propto L$ in the one-boson Krylov subspace $\mathcal{K}_1$ (Cayley tree) (\cref{figS-Lscaling}(a)), and $\langle d\rangle \propto L^3$ in the other three Krylov subspaces (\cref{figS-Lscaling}(c),(e),(g)).

\begin{figure*}[htbp]
\centering
\includegraphics[width=5.4in]{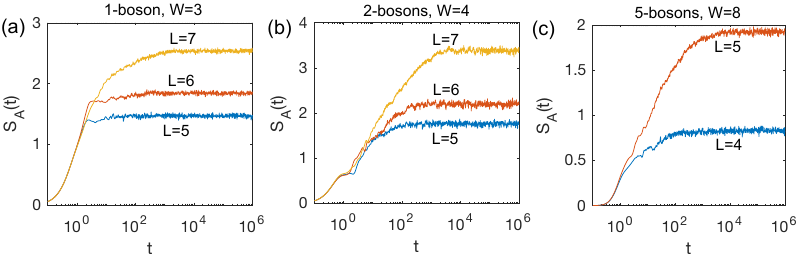}
\caption{time evolution (averaged over many disorder samplings) of entanglement entropy $S_A(t)$ from the root state, for fixed $W$ and different system sizes $L$. (a),(b) and (c) are for the Krylov subspaces (2) (one-boson),(3) (two-boson),(4)(five-boson) in \cref{seq-4Kry}, respectively.}
\label{figS-SAt}
\end{figure*}

We further numerically calculate the root mean square value $|V_\bP|_\text{rms}=\sqrt{\langle V_\bP^2 \rangle }$ of the graph on-site potential $V_\bP=\sum_\bR \epsilon_\bR\langle \rho^\uparrow_\bR\rangle_\bP$, which are shown in \cref{figS-Lscaling}(b),(d),(f),(h) for the four Krylov subspaces in \cref{seq-4Kry}, respectively. The black dashed lines show the values averaged from a large number of disorder samplings, while the red solid line is calculated from the simple rule of square summation:
\begin{equation}
\langle V_\bP^2 \rangle =\sum_\bR \langle \rho^\uparrow_\bR\rangle_\bP \langle \epsilon_\bR^2\rangle =\frac{W^2}{12}\sum_\bR \langle \rho^\uparrow_\bR\rangle_\bP\ ,
\end{equation}
where $\langle \rho^\uparrow_\bR\rangle_\bP=0$ or $1$ if the site has spin $\downarrow$ or $\uparrow$, and $\langle \epsilon_\bR^2\rangle=W^2/12$ is the mean square of $\epsilon_\bR$. The two results (black dashed lines and red solid lines) agree well. The results show $|V_\bP|_\text{rms}\propto L^\gamma W$, with the exponent $\gamma\approx \frac{1}{2}, 1, 1, \frac{1}{2}$ for the four Krylov subspaces in \cref{seq-4Kry}, respectively.

\section{System size scaling of entanglement entropy time evolution}\label{app-EE}

\cref{figS-SAt}(a),(b),(c) show the time evolution (averaged over many disorder samplings) of entanglement entropy $S_A(t)$ with the initial state being the root state, for fixed $W$ and different system sizes $L$, for the Krylov subspaces (2),(3),(4) in \cref{seq-4Kry}, respectively. In all the cases, $S_A(t)\propto \log t$ at early time $t$ and saturates at late time $t$. The time $t_*$ for achieving saturation scales at least as $t_*\sim e^{L/\xi}$ with some $\xi>0$.

\bibliography{cloud_ref}




\end{document}